\def\nuqpo{$\nu_{\rm QPO}$}

\def\nubreak{$\nu_{\rm Break}$} 
\documentstyle[bo99,epsfig]{article}

\def\d705{4U1705-44}
\def\fouru{4U1705-44}
\def\gs{GS1826-238}
\def\nuqpo{$\nu_{\rm QPO}$} 
 
\def\gx{4U1728-34}
\def\tz2{1E1724-3045}

\title{Time Lags in Low Mass X-ray Binaries}

\author{Jean-Fran\c{c}ois Olive \& Didier Barret}                                                       
\affil{Centre d'Etude Spatiale des Rayonnements, CNRS-UPS, 9 Av. du
  Colonel Roche, 31028 Toulouse, Cedex 04, France}                                                
\begin{document}

\maketitle

\begin{abstract}
  Using RXTE/PCA data, we have studied the time lag (TL)
  properties of a sample of four accreting neutron stars (NSs), namely
  \tz2, \gs, \d705~and \gx. The aim of the study is to identify the
  spectral and timing state(s) in which TLs are detected. Along this
  work, we have discovered TLs between the 7-40 keV hard and 2-7 keV
  soft photons from \gx~with amplitudes similar to those seen in
  \fouru~(i.e. $\sim 2$ ms at 5 Hz). We show that the TLs are only
  seen in the low states of those sources, but that within the
  so-called ``island'' spectral state, some sources display TLs
  whereas some do not. On the other hand, we have found that TLs are
  detected when the associated Power Density Spectrum (PDS) shows
  excess power at high frequencies (above $\sim 1$ Hz).
  
\end{abstract}

\section{Introduction}
TLs have been reported so far from both accreting black holes (BHs,
Cyg X-1, GX339-4) and NSs (4U0614+09, 4U1705-44, Ford et al.  1999).
The amplitude of these TLs is typically 20 ms at 1 Hz for NSs and BHs
(Ford et al. 1999).  The origin of these TLs is currently under
debate. Uniform comptonization models predicting constant TLs are
however ruled out by the data.  Comptonization in a non-uniform medium
might account for the observed TLs (Kazanas et al.  1997), but in this
case, their magnitude implies that the Comptonizing cloud extends to
very large radii, which in turn poses an energy problem. Recently,
Poutanen and Fabian (1999) have proposed a ``magnetic flare'' model
that could reproduce the observed PDS and TL magnitude without
requiring a large size region.

Using proprietary and archival RXTE/PCA data, we have initiated a
systematic study of TLs from NSs, with the aim of determining whether
TLs were associated with a singular spectral and/or timing state. We
characterize the spectral state of a source using color-color diagrams
(CCDs). For the timing state, we compute the PDS over a broad X-ray
energy band (2-40 keV). To start with, we have selected two NSs that
are always in a low state (\tz2, \gs, Barret et al. 2000), and two
that are more variable; undergoing occasionally low states (\fouru,
\gx).  For the latter sources, we have sampled both their low and high
states. Their low state PDS are typical of ``island'' state PDS, being
characterized by a flat top below \nubreak~and a broad QPO-like
feature in the declining part of the PDS above \nubreak~(at \nuqpo).
TLs were computed using the techniques described in Nowak et al.
(1999a) between the 7-40 keV hard and 2-7 keV soft photons.

\begin{figure}[!t]
\centerline{\psfig{file=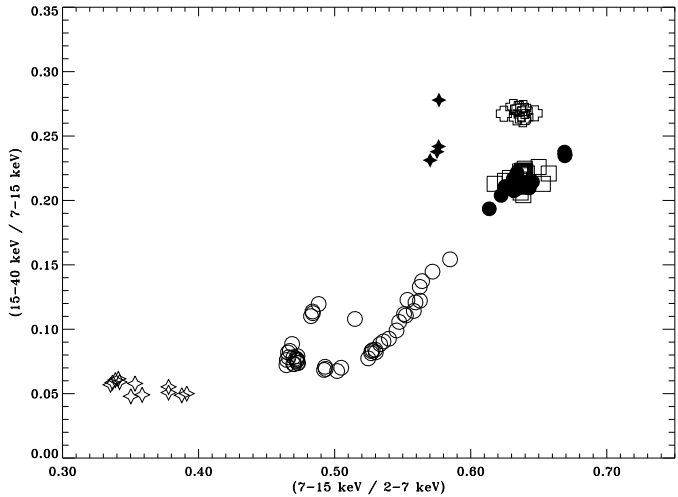,width=8cm,angle=90}}\vspace*{-1.0cm}
\caption[]{Color-Color diagram for \fouru~(stars opened and filled),
  \gx~(circles opened and filled), \tz2~(squares), and \gs~(crosses).
  Filled symbols correspond to time periods during which TLs were
  detected. The integration time of each data point is $\sim$~3000
  seconds. For \gx~and \fouru, both high and low states have been
  sampled. TLs are only detected in their low states (up right in the
  figure).}
\end{figure}

\section{Time Lags versus spectral/timing state}
Figure 1 shows the CCD of the four sources. Fig. 2 shows the
corresponding PDS (top) and TL spectra (bottom). TLs were detected
only from \gx~and \fouru, and in their lowest/hardest intensity states
(namely their ``island'' state). This is the first report of TLs from
\gx. No TLs were detected in their high states, with upper limits of
0.1-0.01 seconds between 1 and 10 Hz; i.e. a factor of 10 larger than
the values detected during their low states. No TLs were detected from
the two steady low state sources, and the upper limits we derived are
lower than the observed values for \gx~and \fouru, indicating that the
non detection of similar TLs is not due to a lack of sensitivity.

\begin{figure}[!t]
\vspace*{-1.0cm}
\centerline{\psfig{file=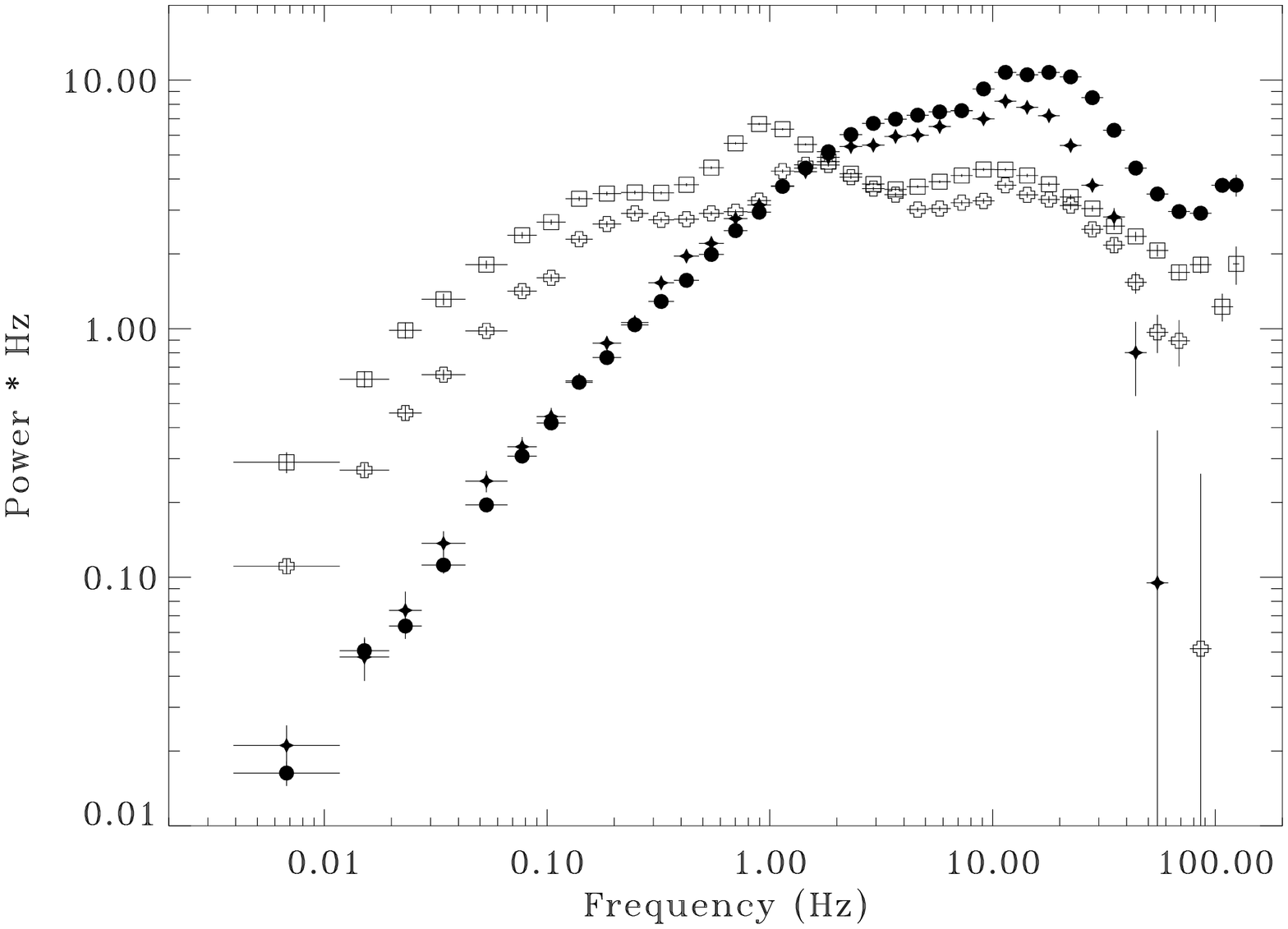, width=7.7cm,
    angle=90}\vspace*{-1.15cm}}
\centerline{\psfig{file=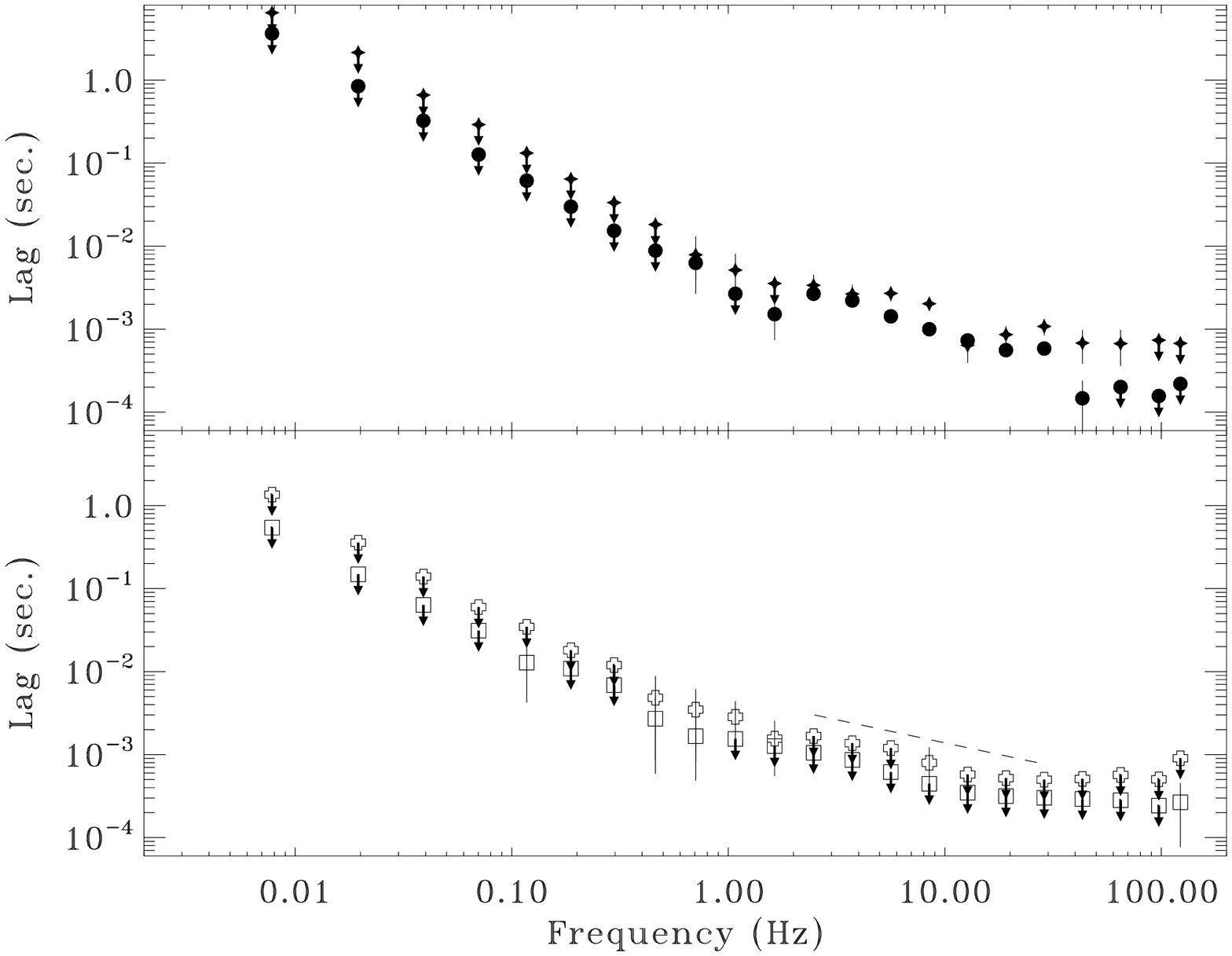, width=7.7cm, angle=90}\vspace*{-.5cm}}
\caption[]{{\it Top panel:~} Power Density Spectra, {\it Bottom panel:~} TL spectra for
  \gx~and \fouru~(top) and upper limits on TLs for \gs~and \tz2~(bottom).
  For indication, the magnitude of the TLs detected from \gx~and
  \fouru~is plotted with a dashed line.}
\end{figure}

Looking at Fig. 1 and 2, a few points can be drawn. First, TLs are not
associated with a singular spectral state; \gx~and \tz2~occupy the
same region of the CCD, and only the former shows TLs. Second, although
the overall shape of their PDS is broadly similar, there is one
noticeable difference that shows up very clearly in the $\nu F \nu$
representation of the PDSs; that TLs are associated with a timing
state in which the whole PDS is shifted towards high frequencies
(\nubreak~and \nuqpo~are a factor of 10 larger for \gx~and \fouru~than
for \tz2~and \gs). Third, when \nubreak~and \nuqpo~are high, TLs are
significantly detected at frequencies between \nubreak~and up to or
slightly above \nuqpo. Fourth, although of lower significance than the
effect observed in the two BHs Cyg X-1 and GX339-4 (Nowak et al.
1999a,b), there is an indication that the TL decreases with
frequency, especially for \gx.  Finally, TLs do not depend upon the
intensity of the aperiodic variability, as the four PDSs of Fig 2.
have comparable integrated RMS (ranging from 17 to 25\% in the 2-40
keV band).

\section{Conclusions}

We have discovered TLs in the low state of \gx, and confirmed the
previous detection of TLs from \fouru~(Ford et al.  1999). In our
attempt to associate the presence of TLs with a singular spectral
state, we have shown that for the same spectral state (same position
in the CCD) some sources do have TLs whereas some do not. The presence
of TLs must thus be tracked somewhere else.

Despite having a limited sample, we have found that TLs are detected
when the characteristic low state PDS shows excess power at high
frequencies (with a \nubreak~$\sim 1$ Hz instead of $\sim 0.1-0.2$
Hz).  Recently, Pottschmidt et al. (2000) showed that the shot
relaxation time in the hard state of Cyg X-1 (scaling as
\nubreak$^{-1}$) anticorrelates with the TL amplitude. We cannot test
the presence of this effect within the data set used here.  However,
if the same anticorrelation applies to NSs, it could provide an
explanation for the non detection of smaller TLs from the sources with
the lowest \nubreak. If \nubreak~and \nuqpo~are somehow related to the
position of the inner disk radius within the corona, as possibly
suggested by recent observations (Revnivtsev et al. 2000), the closer
the disk gets to the central object (i.e the deeper inside the
corona), and the larger the TLs are. In this picture, it is
interesting to note that this subtle change in the accretion geometry
is not reflected in the CCD, implying similar parameters for the
Comptonizing cloud and input seed photon energy for all sources.

To conclude, the above study, although not addressing the origin of
the TL itself has provided some hints about the conditions under which
they are produced. Further work is needed to determine whether TLs are
indeed always associated with the peculiar timing state described in
this paper. Confirming this result would be extremely valuable for
constraining any theoretical models attempting to reproduce the TL
properties of these systems.

We thank E. Ford and A. Zdziarski for helpful comments on this paper.
               
\end{document}